\documentclass[showpacs,preprintnumbers,amsmath,amssymb,onecolumn,floatfix,aps,preprint]{revtex4}

\usepackage{graphicx}% Include figure files
\usepackage{dcolumn}
\usepackage{bm}
\usepackage{epsfig}
\usepackage{amsmath}
\usepackage{amssymb}
\newcommand{\ensemblemean}[1]{\langle\rangle}

\newcommand{\qed}{\nobreak \ifvmode \relax \else
      \ifdim\lastskip<1.5em \hskip-\lastskip
      \hskip1.5em plus0em minus0.5em \fi \nobreak
      \vrule height0.75em width0.5em depth0.25em\fi}

\begin{document}

\title{Feigenbaum graphs: a complex network perspective of chaos}
% Force line breaks with \\

\author{Bartolo Luque$^1$, Lucas Lacasa$^{1*}$, Fernando J. Ballesteros$^2$, Alberto Robledo$^{3,4}$}
\affiliation{$^1$Dept. Matem\'{a}tica Aplicada y Estad\'{i}stica\\
                ETSI Aeron\'{a}uticos, Universidad Polit\'{e}cnica de Madrid, Spain\\
$^2$  Observatori Astron\`{o}mic,\\
                Universitat de Val\`{e}ncia, Spain\\
$^{3}$ Dept. Matem\'{a}ticas, Universidad Carlos III de Madrid, Spain\\
$^{4}$ Instituto de F\'{\i}sica, Universidad Nacional Aut\'{o}noma de M\'{e}xico, Mexico (permanent address)\\
}%
%\author{Visibility team}
%\affiliation{$^1$Departamento de Matem\'atica Aplicada y
%Estad\'istica, ETSI Aeron\'auticos, Universidad Polit\'ecnica de
%Madrid, Madrid 28040, Spain\\\\$^2$Universidad de Valencia }

\date{\today}
% It is always \today, today, % but any date may be
% explicitly specified

\email{lucas.lacasa@upm.es}

%\maketitle%%%

\begin{abstract}

The recently formulated theory of horizontal visibility graphs
transforms time series into graphs and allows the possibility of
studying dynamical systems through the characterization of their
associated networks. This method leads to a natural
graph-theoretical description of nonlinear systems with qualities
in the spirit of symbolic dynamics. We support our claim via the
case study of the period-doubling and band-splitting attractor
cascades that characterize unimodal maps. We provide a universal
analytical description of this classic scenario in terms of the
horizontal visibility graphs associated with the dynamics within
the attractors, that we call Feigenbaum graphs, independent of map
nonlinearity or other particulars. We derive exact results for
their degree distribution and related quantities, recast them in
the context of the renormalization group and find that its fixed
points coincide with those of network entropy optimization.
Furthermore, we show that the network entropy mimics the Lyapunov
exponent of the map independently of its sign, hinting at a
Pesin-like relation equally valid out of chaos.

\end{abstract}

%05.45.Tp   Time series analysis
%05.45.Ac   Low-dimensional chaos
%89.75.Hc   Networks and genealogical trees

\maketitle \noindent\textbf{Introduction\\} We expose a remarkable
relationship between nonlinear dynamical systems and complex
networks by means of the horizontal visibility (HV) algorithm
\cite{pre, pnas, simone} that transforms time series into graphs.
%Specifically, the universal features of the routes to chaos appear associated to the topological and entropic properties of self-similar networks.
In low-dimensional dissipative systems chaotic motion develops out of regular motion in a small number of ways or routes, and amongst
which the period-doubling bifurcation cascade or Feigenbaum scenario is perhaps the better known and most famous mechanism \cite{chaos, chaos2}.
This route to chaos appears an infinite number of times amongst the family of attractors spawned by unimodal maps within the so-called periodic
windows that interrupt stretches of chaotic attractors. In the opposite direction, a route out of chaos accompanies each period-doubling cascade
by a chaotic band-splitting cascade, and their shared bifurcation accumulation points form transitions between order and chaos that are known
to possess universal properties \cite{chaos, chaos2, steve}. Low-dimensional maps have been extensively studied from a purely theoretical
perspective, but systems with many degrees of freedom used to study diverse problems in physics, biology, chemistry, engineering, and
social science, are known to display low-dimensional dynamics \cite{Strogatz}.\\

The horizontal visibility (HV) algorithm converts the information
stored in a time series into a network, setting the nature of the
dynamical system into a different context that requires complex
network tools \cite{stro1, redes, redes2, redes3, bollobas} to
extract its properties. This approach belongs to an emerging
corpus of methods that map series to networks (see for instance
\cite{zhang1,small,pnas,chinos,jstat,thurner} or a recent review
\cite{review_grafos}). Relevant information can be obtained
through the family of visibility methods, including the
characterization of fractal behavior \cite{epl} or the
discrimination between random and chaotic series
\cite{pre,submitted}, and it finds increasing applications in
separate fields, from geophysics \cite{hurricane}, to finance
\cite{finance} or physiology \cite{physiology}. Here we offer a
distinct view of the Feigenbaum scenario through the specific HV
formalism, and provide a complete set of graphs, which we call
Feigenbaum graphs, that encode the dynamics of all stationary
trajectories of unimodal maps. We first characterize their
topology via the order-of-visit and self-affinity properties of
the maps.
%Specifically, we provide a complete analytic characterization of the Feigenbaum scenario, in which expressions for the degree distribution and %related quantities are derived via order-of-visit and self-affinity rules.
Additionally, a matching renormalization group (RG) procedure leads, via its flows, to or from network fixed-points to a comprehensive view of the entire family of attractors. Furthermore, the optimization of the entropy obtained from the degree distribution coincides with the RG fixed points and reproduces the essential features of the map's Lyapunov exponent independently of its sign. A general observation is that the HV algorithm extracts only universal elements of the dynamics, free of the peculiarities of the individual unimodal map, but also of universality classes characterized by the degree of nonlinearity. Therefore all the results presented in this work, while referring to the specific Logistic map for illustrative reasons apply to any unimodal map.\\
%
% This file is part of the APS files in the REVTeX 4 distribution.
% Version 4.0 of REVTeX, August 2001
%
% Copyright (c) 2001 The American Physical Society.
%
% See the REVTeX 4 README file for restrictions and more information.
%
% TeX'ing this file requires that you have AMS-LaTeX 2.0 installed
% as well as the rest of the prerequisites for REVTeX 4.0
%
% See the REVTeX 4 README file
% It also requires running BibTeX. The commands are as follows:
%
% 1) latex apssamp.tex
% 2) bibtex apssamp
% 3) latex apssamp.tex

\begin{figure}[!htb]
\centering
\includegraphics[width=0.8\textwidth]{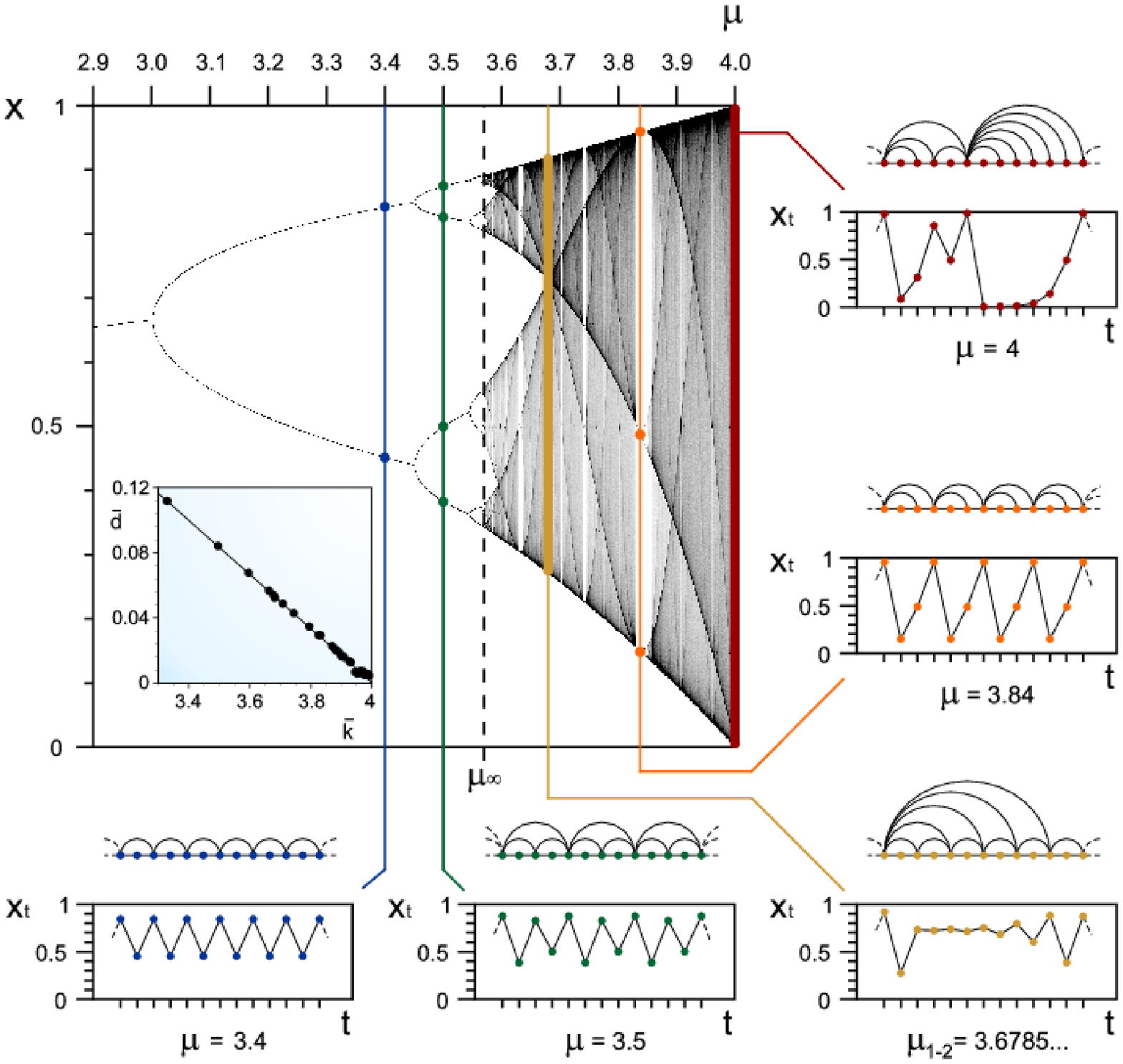}
\caption{\textbf{Feigenbaum graphs from the Logistic map $x_{t + 1}  = f(x_t)=\mu x_t (1 - x_t )$.} The main figure portrays the family of attractors of the Logistic map and indicates a transition from periodic to chaotic behavior at $\mu_\infty = 3.569946...$ through period-doubling bifurcations. For $\mu\geq\mu_\infty$ the figure shows merging of chaotic-band attractors where
aperiodic behavior appears interrupted by windows that, when entered from their left-hand side, display periodic motion of period $T=m\cdot2^0$ with $m>1$ (for $\mu<\mu_{\infty}$, $m=1$) that subsequently develops into $m$ period-doubling cascades with new accumulation points $\mu_\infty(m)$. Each accumulation point $\mu_\infty(m)$ is in turn the limit of a chaotic-band reverse bifurcation cascade with $m$ initial chaotic bands, reminiscent of the self-affine structure of the entire diagram.
All unimodal maps exhibit a period-doubling route to chaos with universal asymptotic scaling ratios between successive bifurcations that depend only on the order of the nonlinearity of the map \cite{Feigenbaum}, the Logistic map belongs to the quadratic case. Adjoining the main figure, we show time series and their associated Feigenbaum graphs according to the HV mapping criterion for several values of $\mu$ where the map evidences both regular and chaotic behavior (see the text). \textit{Inset:} Numerical values of the mean normalized distance $\bar d$ as a function of mean degree $\bar k$ of the Feigenbaum graphs for $3<\mu<4$ (associated to time series of $1500$ data after a transient and a step $\delta \mu=0.05$), in good agreement with the theoretical linear relation (see the text).}
\label{figintro}
\end{figure}

\noindent\textbf{Model: Feigenbaum graphs}\\
The HV graph \cite{pre} associated with a given time series
$\{x_i\}_{i=1,. . .,N}$ of $N$ real data is constructed as
follows: First, a node $i$ is assigned to each datum $x_i$, and
then two nodes $i$ and $j$ are connected if the corresponding data
fulfill the criterion $x_i,x_j > x_n$ for all $n$  such that $i <
n < j$. Let us now focus on the Logistic map \cite{chaos} defined
by the quadratic difference equation $x_{t + 1}  = f(x_t)=\mu x_t
(1 - x_t )$ where $x_t \in [0,1]$ and the control parameter $\mu
\in {\rm [0,4]}$. According to the HV algorithm, a time series
generated by the Logistic map for a specific value of $\mu$ (after
an initial transient of approach to the attractor) is converted
into a Feigenbaum graph (see figure \ref{figintro}). Notice that
this is a well-defined subclass of HV graphs where consecutive
nodes of degree $k=2$, that is, consecutive data with the same
value, do not appear, what is actually the case for series
extracted from maps (besides the trivial case of a constant
series). While for a period $T$ there are in principle several
possible periodic orbits, and therefore the set of associated
Feigenbaum graphs is degenerate, it can be proved that the mean
degree $\bar k(T)$ and normalized mean distance $\bar d(T)$ of all
these Feigenbaum graphs fulfill $\bar k(T)=4(1-\frac{1}{2T})$ and
$\bar d(T)=\frac{1}{3T}$ respectively, yielding a linear relation
$\bar d(\bar k)=(4-\bar k)/6$ that is corroborated in the inset of
figure \ref{figintro}. Observe that aperiodic series ($T\rightarrow\infty$) reach the upper bound mean degree $\bar k=4$.\\

\begin{figure}[!htb]
\centering
\includegraphics[width=0.8\textwidth]{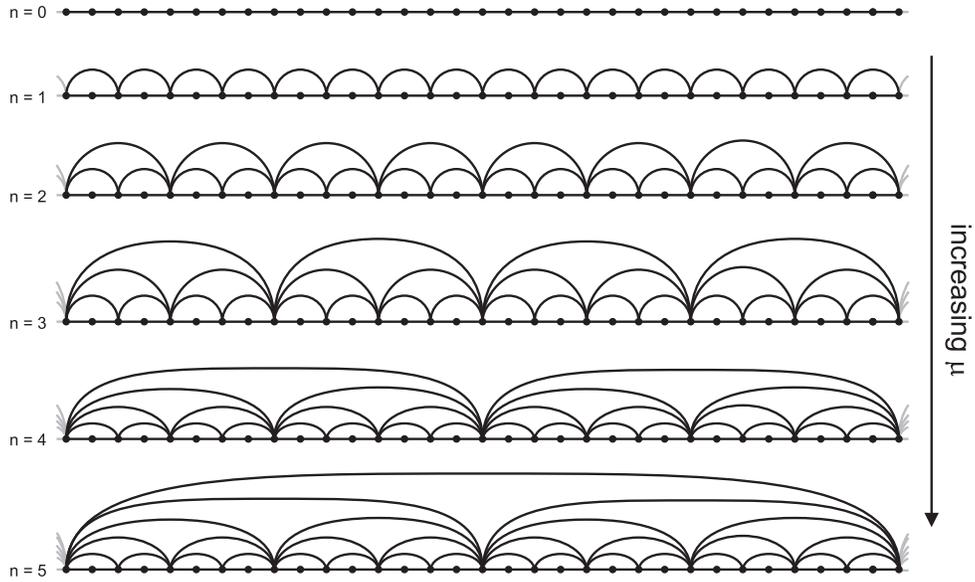}
\caption{\textbf{Periodic Feigenbaum graphs for $\mu<\mu_\infty$.} The sequence of graphs associated to periodic attractors with increasing period $T=2^n$ undergoing a period-doubling cascade. The pattern that occurs for increasing values of the period is related to the universal ordering with which an orbit visits the points of the attractor. Observe that the hierarchical self-similarity of these graphs requires that the graph for $n-1$ is a subgraph of that for $n$.}
\label{grafos de feigenbaum1}
\end{figure}

\noindent\textbf{Results\\} A deep-seated feature of the
period-doubling cascade is that the order in which the positions
of a periodic attractor are visited is universal \cite{schroeder},
the same for all unimodal maps. This ordering turns out to be a
decisive property in the derivation of the structure of the
Feigenbaum graphs. See figure \ref{grafos de feigenbaum1} where we
plot the graphs for a family of attractors of increasing period
$T=2^n$, that is, for increasing values of $\mu<\mu_{\infty}$.
This basic pattern also leads to the expression for their
associated degree distributions,

\begin{eqnarray}\label{degree_p_1_n}
& P(n,k) = \left( {\frac{1}{2}} \right)^{k/2}, \quad k = 2, 4, 6, ...,2n,\\
& P(n,k) = \left( {\frac{1}{2}} \right)^{n}, \quad k = 2(n+1),\nonumber
\end{eqnarray}
and zero for $k$ odd or $k>2(n+1)$. At the accumulation point
$\mu_{\infty}$ the period diverges ($n \rightarrow \infty$) and
the distribution is exponential for all even values of the degree,
\begin{equation}\label{degree_p_1_critico}
P(\infty,k) = \left(\frac{1}{2}\right)^{k/2}, \quad  k = 2, 4, 6, ...,
\end{equation}
and zero for $k$ odd. Observe that these relations are independent of the
order of the map's nonlinearity: the HV algorithm sifts out every detail of the dynamics except for the basic storyline.\\

\begin{figure}[!htb]
\centering
\includegraphics[width=0.8\textwidth]{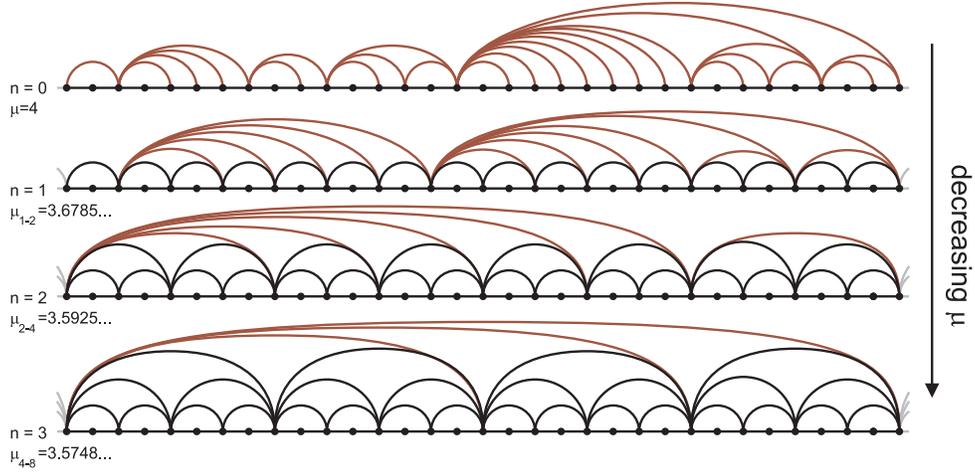}
\caption{\textbf{Aperiodic Feigenbaum graphs for $\mu>\mu_\infty$.} A sequence of graphs associated with chaotic series after $n$ chaotic-band reverse bifurcations, starting at $\mu=4$ for $n=0$, when the attractor extends along a single band and the degree distribution does not present any regularity (red links). For $n>0$ the phase space is partitioned in $2^n$ disconnected chaotic bands and the $n$-th self-affine image of $\mu=4$ is the $n$-th Misiurewicz point $\mu_{2^{n-1}-2^n}$. In all cases, the orbit visits each chaotic band in the same order as in the periodic region $\mu<\mu_{\infty}$. This order of visits induces an ordered structure in the graphs (black links) analogous to that found for the period-doubling cascade.}
\label{grafos de feigenbaum2}
\end{figure}

 We turn next to the period-doubling bifurcation cascade of chaotic
 bands that takes place as $\mu$ decreases from $\mu=4$ towards $\mu_{\infty}$.
 For the largest value of the control parameter, at $\mu=4$, the attractor is fully chaotic and
 occupies the entire interval $[0,1]$ (see figure \ref{figintro}). This is the first chaotic
 band $n=0$ at its maximum amplitude. As $\mu$ decreases in value within  $\mu_{\infty}<\mu<4$
 band-narrowing and successive band-splittings \cite{chaos, chaos2, steve, schroeder} occur. In general,
 after $n$ reverse bifurcations the phase space is partitioned in $2^n$ disconnected chaotic bands, which
 are self-affine copies of the first chaotic band \cite{mandelbrot}. The values of $\mu$ at which the
 bands split are called Misiurewicz points \cite{schroeder}, and their location converges to the
 accumulation point $\mu_{\infty}$ for $n\rightarrow\infty$. Significantly, while in the chaotic
 zone orbits are aperiodic, for reasons of continuity they visit each of the $2^n$ chaotic bands
 in the same order as positions are visited in the attractors of period $T=2^n$ \cite{schroeder}.
 In figure \ref{grafos de feigenbaum2} we have plotted the Feigenbaum graphs generated through chaotic
 time series at different values of $\mu$ that correspond to an increasing number of reverse bifurcations.
 Since chaotic bands do not overlap, one can derive the following degree distribution for a Feigenbaum graph
 after $n$ chaotic-band reverse bifurcations by using only the universal order of visits

\begin{eqnarray}\label{degree_p_mu}
& P_{\mu}(n,k)=\left(\frac{1}{2}\right)^{k/2}, \quad k = 2, 4, 6,...,2n, \nonumber\\
& P_{\mu}(n,k\geq 2(n+1))=\left(\frac{1}{2}\right)^n,
\end{eqnarray}

\noindent and zero for $k = 3, 5, 7,...,2n+1$. We note that this time the
degree distribution retains some dependence on the specific value of $\mu$,
concretely, for those nodes with degree $k\geq 2(n+1)$, all of which belong to the
top chaotic band (labelled with red links in figure \ref{grafos de feigenbaum2}). The
HV algorithm filters out chaotic motion within all bands except for that taking place
in the top band whose contribution decreases as $n \to \infty$ and appears coarse-grained
in the cumulative distribution $P_{\mu}(n,k\geq 2(n+1))$. As would be expected, at the
accumulation point $\mu_{\infty}$ we recover the exponential degree distribution (equation \ref{degree_p_1_critico}),
\textit{i.e.} $\lim_{n\rightarrow\infty}P_{\mu}(n,k)=P(\infty,k)$.\\

%as the chaotic nature of the attractor is encoded, by construction, in the connectivity pattern associated to the top chaotic band. This %contribution appears coarse-grained in the cumulative distribution $P_{\mu}(n,k\geq 2(n+1))$. The HV algorithm filters out chaotic motion within all %bands except for that taking place in the top band. The contribution to the degree distribution of the chaotic nature of the top band decreases as %$n \to \infty$, and, as it should be expected, at the accumulation point $\mu_{\infty}$ we recover the exponential degree distribution (equation %\ref{degree_p_1_critico}), \textit{i.e.} $\lim_{n\rightarrow\infty}P_{\mu}(n,k)=P(\infty,k)$.\\

Before proceeding to interpret these findings via the consideration of
renormalization group (RG) arguments, we recall that the Feigenbaum tree
shows a rich self-affine structure: for $\mu>\mu_{\infty}$ periodic windows of
initial period $m$ undergo successive period-doubling bifurcations with
new accumulation points $\mu_{\infty}(m)$ that appear interwoven with chaotic attractors.
These cascades are self-affine copies of the fundamental one. The process of reverse
bifurcations also evidences this self-affine structure, such that each accumulation
point is the limit of a chaotic-band reverse bifurcation cascade. Accordingly,
we label $G(m,n)$ the Feigenbaum graph associated with a periodic series of
period $T=m \cdot 2^n$, that is, a graph obtained from an attractor within
window of initial period $m$ after $n$ period-doubling bifurcations. In
the same fashion, $G_{\mu}(n,m)$ is associated with a chaotic attractor
composed by $m \cdot 2^n$ bands (that is, after $n$ chaotic band reverse bifurcations
of $m$ initial chaotic bands). Therefore, graphs depicted in
figures \ref{grafos de feigenbaum1} and \ref{grafos de feigenbaum2} correspond to $G(1,n)$ and $G_\mu(1,n)$
respectively and for the first accumulation point we have $G(1,\infty)=G_\mu(1,\infty)\equiv G_\infty$. Similarly,
in each accumulation point $\mu_\infty(m)$ we have $G(m,\infty)=G_\mu(m,\infty)$.\\

\begin{figure}[!htb]
\centering
\includegraphics[width=0.9\textwidth]{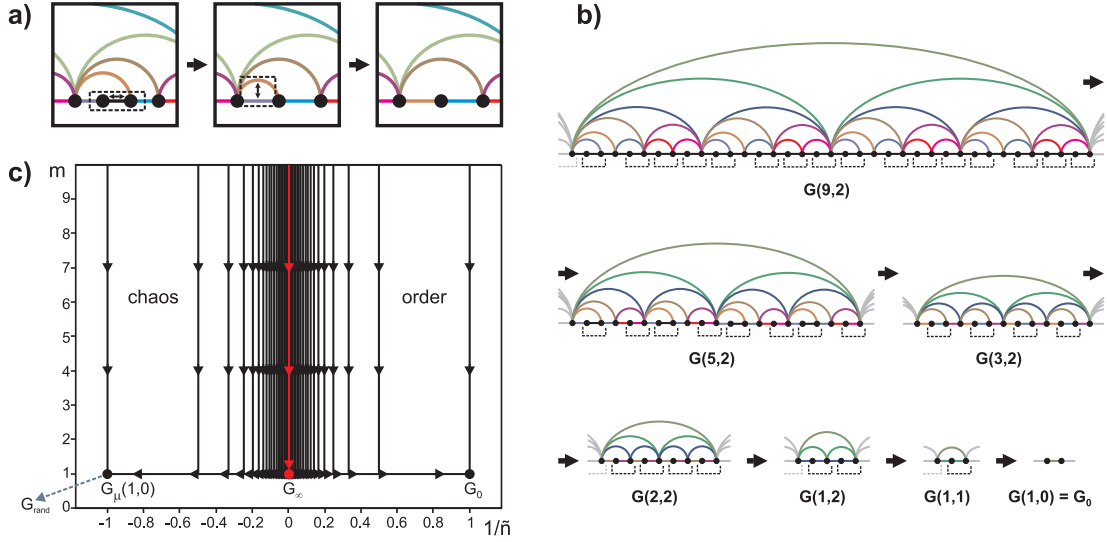}
\caption{\textbf{Renormalization process and network RG flow structure.} (a) Illustration of the renormalization process $\cal R$: a node with degree $k=2$ is coarse-grained with one of its neighbors (indistinctively) into a block node that inherits the links of both nodes. This process coarse-grains every node with degree $k=2$ present at each renormalization step. (b) Example of an iterated renormalization process in a sample Feigenbaum graph at a periodic window with initial period $m=9$ after $n=2$ period-doubling bifurcations (an orbit of period $T=m\cdot2^n=36$).
(c) RG flow diagram, where $m$ identifies the periodic window that is initiated with period $m$ and \textit{\~{n}} designates the order of the bifurcation, \textit{\~{n}} $=n+1$ for period-doubling bifurcations and \textit{\~{n}} $=-(n+1)$ for reverse bifurcations.
$\Delta \mu(m)=\mu_\infty(m)-\mu$ denotes the reduced control parameter of the map, and $\mu_\infty(m)$ is the location of the accumulation point of the bifurcation cascades within that window. Feigenbaum graphs associated with periodic series ($\Delta \mu (m)>0$, \textit{\~{n}} $>0$) converge to $G(1,0)\equiv G_0$ under the RG, whereas those associated with aperiodic ones ($\Delta\mu(m)<0$, \textit{\~{n}} $<0$) converge to $G_{\text{rand}}$. The accumulation point $\mu_\infty\equiv\mu_\infty(1)$ corresponds to the unstable (nontrivial) fixed point $G(1,\infty)\equiv G_\infty$ of the RG flow, which is nonetheless approached through the critical manifold of graphs $G(m,\infty)$ at the accumulation points $\mu_\infty(m)$. In summary, the nontrivial fixed point of the RG flow is only reached via the family of the accumulation points, otherwise the flow converges to trivial fixed points for periodic or chaotic regions.}
\label{RGfig}
\end{figure}

%\noindent\textbf{Renormalization Group}\\
 In order to recast previous findings in the context of the renormalization group, let us define an RG operation $\cal R$ on a graph as the
 coarse-graining of every couple of adjacent nodes where one of them has degree $k=2$ into a block node that inherits the links of the previous
 two nodes (see figure \ref{RGfig}.a). This is a real-space RG transformation on the Feigenbaum graph \cite{newmannRG}, dissimilar from recently
 suggested box-covering complex network renormalization schemes \cite{rg, rg2, rg3}. This scheme turns out to be equivalent for $\mu<\mu_{\infty}$ to the construction of an HV graph from the composed map $f^{(2)}$ instead of the original $f$, in correspondence to the original Feigenbaum renormalization procedure \cite{Feigenbaum, steve}. We first note that ${\cal R}\{G(1,n)\}=G(1,n-1)$, thus, an iteration of this process yields an RG flow that converges to the (1st) trivial fixed point ${\cal R}^{(n)}\{G(1,n)\}= G(1,0)\equiv G_0= {\cal R}\{G_0 \}$. This is the stable fixed point of the RG flow $\forall \mu<\mu_{\infty}$. We note that there is
  only one relevant variable in our RG scheme, represented by the reduced
  control parameter $\Delta \mu=\mu_{\infty}-\mu$, hence, to identify a nontrivial fixed
  point we set $\Delta \mu=0$ or equivalently $n \to \infty$, where the structure of the Feigenbaum graph turns to
  be completely self-similar under $\cal R$. Therefore we conclude that $G(1,\infty)\equiv G_\infty$ is the nontrivial
  fixed point of the RG flow, ${\cal R}\{G_\infty \}=G_\infty$. In connection with this, let $P_t(k)$ be the degree
  distribution of a generic Feigenbaum graph $G_t$ in the period-doubling cascade after $t$ iterations of $\cal R$,
  and point out that the RG operation, ${\cal R}\{G_t\}=G_{t+1}$, implies a recurrence relation $(1-P_t(2))P_{t+1}(k)=P_t(k+2)$,
  whose fixed point coincides with the degree distribution found in equation \ref{degree_p_1_critico}. This confirms that the
  nontrivial fixed point of the flow is indeed $G_\infty$.\\

Next, under the same RG transformation, the self-affine structure of the family of attractors yields ${\cal R}\{G_{\mu}(1,n)\}=G_{\mu}(1,n-1)$,
generating a RG flow that converges to the Feigenbaum graph associated to the 1st chaotic band, ${\cal R}^{(n)}\{G_{\mu}(1,n) \}=G_{\mu}(1,0)$.
Repeated application of $\cal R$ breaks temporal correlations in the series, and the RG flow leads to a 2nd trivial
fixed point ${\cal R}^{(\infty)}\{G_{\mu}(1,0)\}=G_{\text{rand}}={\cal R}\{G_{\text{rand}}\}$, where $G_{\text{rand}}$ is the HV graph generated by a
purely uncorrelated random process. This graph has a universal degree distribution $P(k)=(1/3)(2/3)^{k-2}$, independent of the random process
underlying probability density (see \cite{pre, submitted}).\\

Finally, let us consider the RG flow inside a given periodic window of initial period $m$. As the renormalization process addresses
nodes with degree $k=2$, the initial applications of $\cal R$ only change the core structure of the graph associated with the
specific value $m$ (see figure \ref{RGfig}.b for an illustrative example). The RG flow will therefore converge to the 1st trivial fixed point
via the initial path ${\cal R}^{(p)}\{G(m,n)\}=G(1,n)$, with $p \le m$, whereas it converges to the 2nd trivial fixed point
for $G_{\mu}(m,n)$ via ${\cal R}^{(p)}\{G_{\mu}(m,n)\}=G_{\mu}(1,n)$.  In the limit of $n\rightarrow\infty$ the RG flow proceeds towards the
nontrivial fixed point via the path ${\cal R}^{(p)}\{G(m,\infty)\}=G(1,\infty)$. Incidentally, extending the definition of the reduced control
parameter to $\Delta \mu(m)=\mu_\infty(m)-\mu$, the family of accumulation points is found at $\Delta\mu(m)=0$. A complete schematic representation
of the RG flows can be seen in figure \ref{RGfig}.c.\\

Interestingly, and at odds with standard RG applications to (asymptotically) scale-invariant systems, we find that
invariance at $\Delta \mu=0$ is associated in this instance to an exponential (rather than power-law) function of the observables,
concretely, that for the degree distribution. The reason is straightforward: ${\cal R}$ is not a conformal transformation ($\textit{i.e.}$ a
scale operation) as in the typical RG, but rather, a translation procedure. The associated invariant functions are therefore non
homogeneous (with the property $\text{g}(ax)=b\text{g}(x)$), but exponential (with the property $\text{g}(x+a)=c\text{g}(x)$).\\

%\noindent\textbf{Network entropy}\\
Finally, we derive, via optimization of an entropic functional for
the Feigenbaum graphs, all the RG flow directions and fixed points
directly from the information contained in the degree
distribution. Amongst the graph theoretical entropies that have
been proposed we employ here the Shannon entropy of the degree
distribution $P(k)$, that is $h=-\sum_{k=2}^{\infty}  {P(k)\log
P(k)}$. By making use of the Maximum Entropy formalism, it is easy
to prove that the degree distribution $P(k)$ that maximizes $h$ is
exactly $P(k)=(1/3)(2/3)^{k-2}$, which corresponds to
the distribution for the 2nd trivial fixed point of the RG flow $G_{\text{rand}}$. Alternatively, with the incorporation of the
additional constraint that allows only even values for the degree (the topological restriction for Feigenbaum graphs $G(1,n)$),
entropy maximization yields a degree distribution that coincides with equation \ref{degree_p_1_critico}, which corresponds to the nontrivial
fixed point of the RG flow $G_\infty$. Lastly, the degree distribution that minimizes $h$ trivially corresponds to $G_0$, \textit{i.e.} the
1st trivial fixed point of the RG flow. Remarkably, these results indicate that the fixed-point structure of the RG flow are obtained via
optimization of the entropy for the entire family of networks, supporting a suggested connection between RG theory and the principle of
Maximum Entropy \cite{robledo}.\\

\begin{figure}[!htb]
\centering
\includegraphics[width=0.8\textwidth]{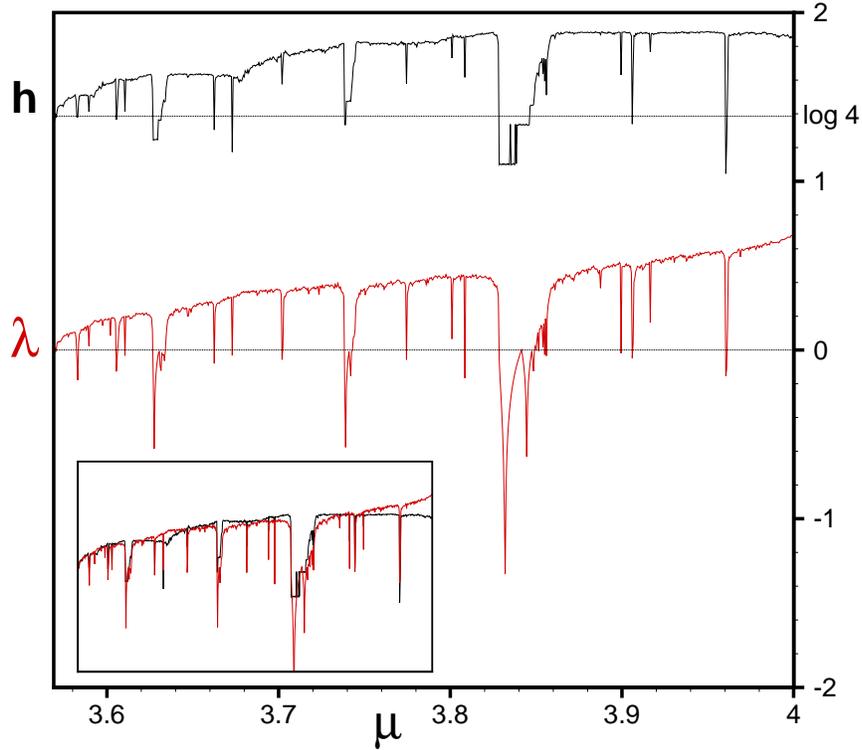}
\caption{\textbf{Horizontal visibility network entropy $h$ and Lyapunov exponent $\lambda$ for the Logistic map.} We plot the numerical values of $h$ and $\lambda$ for $3.5<\mu<4$ (the numerical step is $\delta\mu=5\cdot10^{-4}$ and in each case the processed time series have a size of $2^{12}$ data). The inset reproduces the same data but with a rescaled entropy $h-\log(4)$.  The surprisingly good match between both quantities is reminiscent of the Pesin identity (see text). Unexpectedly, the Lyapunov exponent within the periodic windows ($\lambda<0$ inside the chaotic region) is also well captured by $h$.} \label{entropy}
\end{figure}

\noindent The network entropy $h$ can be calculated exactly for
$G(1,n)$ ($\mu<\mu_\infty$ or $T=2^n$), yielding $h(n)=\log
4\cdot(1-{2^{-n}})$. Because increments of entropy are only due to
the occurrence of bifurcations $h$ increases with $\mu$ in a
step-wise way, and reaches asymptotically the value
$h(\infty)=\log 4$ at the accumulation point $\mu_{\infty}$. For
Feigenbaum graphs $G_\mu(1,n)$ (in the chaotic region), in general
$h$ cannot be derived exactly since the precise shape of $P(k)$ is
unknown (albeit the asymptotic shape is also exponential
\cite{submitted}). Yet, the main feature of $h$ can be determined
along the chaotic-band splitting process, as each reverse
bifurcation generates two self-affine copies of each chaotic band.
Accordingly, the decrease of entropy associated with this reverse
bifurcation process can be described as $h_\mu(n)=\log 4
+h_\mu(0)/{2^{n}}$, where the entropy $h_\mu(n)$ after $n$ reverse
bifurcations can be described in terms of the entropy associated
with the first chaotic band $h_\mu(0)$. In figure \ref{figintro}
we observe how the chaotic-band reverse bifurcation process takes
place in the chaotic region from right to left, and therefore
leads in this case to a decrease of entropy with an asymptotic
value of $\log 4$ for $n\rightarrow\infty$ at the accumulation
point. These results suggest that the graph entropy behaves
qualitatively as the map's Lyapunov exponent $\lambda$, with the
peculiarity of having a shift of  $\log 4$, as confirmed in figure
\ref{entropy}.
 This unexpected qualitative agreement is reasonable in the chaotic region in view of the
 Pesin theorem \cite{chaos2}, that relates the positive Lyapunov exponents of a map with its Kolmogorov-Sinai entropy
 (akin to a topological entropy) that for unimodal maps reads $h_{KS}=\lambda,\ \forall \lambda>0$, since $h$ can
 be understood as a proxy for $h_{KS}$. Unexpectedly, this qualitative agreement seems also valid in the periodic
 windows ($\lambda<0$), since the graph entropy is positive and varies with the
 value of the associated (negative) Lyapunov exponent even though $h_{KS}=0$, hinting at a Pesin-like relation valid also out of chaos
 which deserves further
 investigation.
 The agreement between both quantities lead us to conclude that the Feigenbaum graphs capture not only the period-doubling
 route to chaos in a universal way, but also inherits the main feature of chaos, \textit{i.e.} sensitivity to initial conditions.\\

\noindent\textbf{Discussion\\} In summary, we have shown that the
horizontal visibility theory combines power with
straightforwardness as a tool for the analytical study of
nonlinear dynamics. As an illustration we have established how the
families of periodic and chaotic attractor bifurcation cascades of
unimodal maps transform into families of networks with
scale-invariant limiting forms, whose characterization can be
deduced from two basic and universal properties of unimodal maps:
ordering of consecutive positions in the attractors and
self-affinity. Further, we have demonstrated that these networks
and their associated degree distributions comply with
renormalization group and maximum entropy principles, filtering
out irrelevant variables and finding fixed-point networks which
are independent of the map's nonlinearity. The entire Feigenbaum
scenario is therefore fully described. The potential of the theory
for revealing new information is indicated by the ability of the
network entropy to emulate the Lyapunov exponent for both periodic
and chaotic attractors. Extensions of this approach to other
complex behavior, such as dynamical complexity associated to
vanishing Lyapunov exponents, intermittency, quasiperiodic routes
to chaos, etc., are still open questions. Finally, observe that in
symbolic dynamics \cite{hao} one usually defines a phase-space
partition (Markov partition) in order to create a symbolic
representation of the dynamics. This partition tiles phase space
in a non-overlapping manner: every value of the series has a
univocally associated symbol. While a Feigenbaum graph also
symbolizes the series data (incidentally, without the need of
defining an ad hoc partition), each series datum is not associated
univocally to a symbol (the degree of the node): this symbol is a
function, in principle, of the complete series, and incorporates
global information. Furthermore, note that besides the
symbolization that converts a time series into a series of node
degrees, the Feigenbaum graphs also store the connectivity pattern
amongst nodes -i.e. the topological structure of the graph. On
this respect, the possible connections of HV theory with kneading
theory \cite{kneading} and symbolic dynamics \cite{hao} are of
special interest. \\\\

\noindent \textbf{Acknowledgments}\\
The authors acknowledge comments from anonymous referees.
%BL and
%LL acknowledge financial support from the MEC and Comunidad de
%Madrid (Spain) through projects FIS2009-13690 and S2009ESP-1691.
%FJB acknowledges support from MEC through project AYA2006-14056,
%and AR acknowledges support from MEC (Spain) and CONACyT and
%DGAPA-UNAM (Mexican agencies).

%\bibliography{apssamp}% Produces the bibliography via BibTeX.

\newpage

\noindent\textbf{Figure Legends}

\end{document}